# Highly Quantum-Confined InAs Nanoscale Membranes

Kuniharu Takei[1,2,3,†], Hui Fang[1,2,3,†], Bala Kumar[4,†], Rehan Kapadia[1,2,3], Qun Gao[4], Morten Madsen[1,2,3], Ha Sul Kim[1,2,3], Chin-Hung Liu[5], Yu-Lun Chueh[5], Elena Plis[6], Sanjay Krishna[6], Hans A. Bechtel[7], Jing Guo[4], Ali Javey[1,2,3,*]

[1]Electrical Engineering and Computer Sciences, University of California, Berkeley, CA, 94720.
[2]Materials Sciences Division, Lawrence Berkeley National Laboratory, Berkeley, CA 94720.
[3]Berkeley Sensor and Actuator Center, University of California, Berkeley, CA, 94720.
[4]Electrical and Computer Engineering, University of Florida, Gainsville, FL, 32611
[5]Materials Science and Engineering, National Tsing Hua University, Hsinchu 30013, Taiwan, R.O.C.
[6]Center for High Technology Materials, University of New Mexico, Albuquerque, NM 87106.
[7]Advanced Light Source, Lawrence Berkeley National Laboratory, Berkeley, CA 94720

[†] These authors contributed equally to this work.
* Correspondence should be addressed to A.J. (ajavey@eecs.berkeley.edu).

**Abstract -** Nanoscale size-effects drastically alter the fundamental properties of semiconductors. Here, we investigate the dominant role of quantum confinement in the field-effect device properties of free-standing InAs nanomembranes with varied thicknesses of 5-50 nm. First, optical absorption studies are performed by transferring InAs "quantum membranes" (QMs) onto transparent substrates, from which the quantized sub-bands are directly visualized. These sub-bands determine the contact resistance of the system with the experimental values consistent with the expected number of quantum transport modes available for a given thickness. Finally, the effective electron mobility of InAs QMs is shown to exhibit anomalous field- and thickness-dependences that are in distinct contrast to the conventional MOSFET models, arising from the strong quantum confinement of carriers. The results provide an important advance towards establishing the fundamental device physics of 2-D semiconductors.

2-D materials are of profound interest for exploring fundamental properties at the nanoscale[1,2,3] and enabling novel practical applications, such as high performance electronics[4,5,6] or quantum optics[7]. Over the past several years, various layered semiconductors, including graphene and MoS$_2$, have been extensively studied owing to their intriguing properties[6,8,9,10,11]. The InAs QMs reported here present another class of 2-D materials, where the band-structure can be precisely tuned from bulk to 2-D by changing the thickness. Specifically, given the large Bohr radius of ~34 nm for bulk InAs, a strong quantum confinement is expected, even for layers with 10's of nm in thickness. Thus, the basic material properties can be systematically studied as the material is quantum confined with the size effects being fundamentally different than those of the layered semiconductors. In contrast to conventional III-V quantum-well (QW) structures where confinement is achieved by the use of a large band-gap semiconductor layer on an epitaxial growth substrate[12,13], QMs are confined at the surface by either an insulator layer or air/vacuum[15,14], and are free to be placed on a user-defined substrate. This QM structural configuration enables the direct contact of the active semiconductor layer with the gate stack, which is essential for enabling high performance devices. We recently demonstrated the promise of this material concept by the fabrication of ultrathin-body InAs transistors on Si/SiO$_2$ support substrates with excellent electrical properties[15], taking advantage of the high electron mobility of III-V semiconductors[16,17,18] and the well-established processing technology of Si. Here, we use InAs QMs as a model system for probing the drastic role of quantum confinement on the contact resistance and carrier transport properties of field-effect transistors (FETs) as a function of material thickness through detailed experiments and theoretical modeling. The work presents fundamental understanding of the device physics and the projected performance limits of nanoscale III-V FETs.

InAs QMs were fabricated on Si/SiO$_2$ and CaF$_2$ substrates by using an epitaxial layer transfer (ELT) method[15,19,20,21] (see Methods for process detail). Figures 1a & b show the low and high resolution transmission electron microscopy (TEM) images of a typical 7 nm InAs membrane capped by SiO$_2$ on a Si/SiO$_2$ substrate, illustrating the high quality, single-crystalline InAs QM of just a few atomic layers in thickness. The interfaces between InAs and SiO$_2$ are abrupt as evident from TEM. Fourier transform infrared (FTIR) spectroscopy was used to directly probe the optical absorption transitions of InAs QMs, thereby experimentally elucidating the electronic band structure quantization as a function of InAs thickness. Here, InAs QMs were transferred onto a CaF$_2$ substrate, which is transparent for the photon wavelengths of interest (1-4 μm). Figure 2a shows the normalized transmittance spectra of 6-19 nm thick InAs QMs at room temperature. The clear steps observed in the transmittance are attributed to the interband transitions between the 2-D sub-bands, with the spacing of the steps decreasing as the QM thickness is increased.

The onset energies of the absorption steps can be estimated by $E=E_g+E_{hn}+E_{en'}$, where $E_g$ is the band gap of bulk InAs, and $E_{hn}$ ($E_{en'}$) are the n (n') -th sub-band energy of holes (electrons) due to quantum confinement. Note that due to the band edge roughness induced by various surface defects (e.g. interface traps, surface roughness, etc.), the absorption edge is not strictly sharp. By modeling InAs QM as a finite potential well structure, the n'th sub-band energy can be numerically calculated from the 1-D Schrodinger equation (see Supp. Info. for calculation details), as shown in Fig. 2b. Selection rules require the sub-bands involved in the transition to have matching envelope functions, or parity. Consequently, electrons in the first electron sub-band (e1) can only transit to the first hole sub-bands (either 1$^{st}$ heavy hole, hh1; light hole, lh1; or split-off, so1), i.e., Δn=n'-n=0. Other transitions are also possible but are weak.

Figure 2c shows the FTIR absorption onsets for each QM thickness (see Supp. Info., Fig. S2), along with the calculated interband transition energies (e1hh1, e1lh1, e2hh2, e2lh2). The experimental results closely match the theoretical calculations. The FTIR data is a direct observation of the electronic band structure quantization in ultrathin InAs QMs. This clearly presents an important utility of the proposed platform for exploring the basic properties of nanoscale material by decoupling the active material from the original growth (i.e., GaSb) substrate.

Next, the effect of quantum confinement on the contact resistance, $R_c$, of InAs QMs is explored by fabricating back-gated, field-effect transistors on Si/SiO$_2$ substrates. Ni source/drain (S/D) metal contacts were used for this study as Ni forms ohmic contact to the conduction band of InAs, with negligible parasitic junction resistances (i.e., without tunneling barriers) for electron injection[22]. This enables the study of the fundamental limits of the metal-semiconductor junction resistance of QMs as a function of thickness. The transmission line method (TLM) [10] was used to extract the contact resistance. Typical $I_{DS}$-$V_{GS}$ curves of a back-gated (50 nm SiO$_2$ gate dielectric) 18 nm-thick InAs QM with different channel lengths are shown in Fig. 3a. The ON-state resistance at $V_{GS}$-$V_{th}$=9 V, where $V_{th}$ is the threshold voltage, as a function of $L$ (measured by SEM) was extracted for InAs QMs with different thicknesses and is plotted in Fig. 3b. The ON-state resistance monotonically increases with $L$ since the devices are operating in the diffusive regime for the explored length scales. The y-intercept of the ON-resistance versus $L$ plot is approximately equal to $2R_c$. The extracted contact resistance is plotted as a function of InAs thickness in Fig. 3c. A low contact resistance value of ~80 Ω-μm is observed for $T_{InAs} \geq 10$ nm. As the thickness is reduced to $\leq 10$ nm, a sharp increase in the resistance is observed with $R_c$~ 600 Ω-μm at $T_{InAs} = 5$ nm.

In the ideal limit (i.e., in the absence of parasitic resistances), the contact resistance of a QM approaches the quantum resistance, $R_Q$, which is a function of the number of accessible transport modes (Fig. S5) as given by the following analytical expression[23],

$$R_Q = \frac{h}{M 2e^2} = \frac{h}{2e^2} \sum_{i=1}^{N} \frac{1}{Int\left(\frac{Wk_{Fi}}{\pi}\right)}$$

Here, index $i$ labels the sub-bands, $k_{Fi}$ is the Fermi wave vector of the $i^{th}$ sub-band, $N$ is the highest filled sub-band, $M$ is the number of available transport modes, $h/2e^2$ is the quantum of resistance, and $W$ is the width of the system. In this analysis, each 2-D sub-band can be considered to be composed of multiple parallel 1-D sub-bands or transport modes, with the resistance of a single mode being $h/2e^2$. The calculated $R_Q$ (see Supp. Info. for calculation details) assumes that all modes with energy lower than the Fermi level contribute to current. Since this analysis ignores the effects of the transmission coefficient at the contact interfaces, the calculated $R_Q$ only qualitatively matches the experimental results (Fig. 3c).

To account for the barrier height between the metal contact and individual sub-bands in the QM, a non-equilibrium Green's function (NEGF) calculation method was used[24]. The simulation models the quantum transport from the 3-D metal/InAs contact to the 2-D InAs QM (see Supp. Info. for simulation details). The simulation indicates that the InAs regions underneath the metal contacts are effectively a 3-D material system with the density of states not exhibiting quantized sub-bands. The lack of significant quantum confinement in the contact regions arises from the barrier-free contact for electron injection. Here, we assumed a negative Schottky barrier height of -0.15 eV between Ni and InAs conduction band edge, and a uniform charge density of $3\times10^{18}$ cm$^{-3}$ in the membrane which is the surface electron concentration previously reported for InAs[25,26]. The experimental and ballistic NEGF simulation curves are nearly identical (Fig. 3c), providing a strong evidence that the measured $R_c$ purely arises from the quantum resistance of the system and the associated transmission probability for each transport mode, without parasitic resistances.

Quantization is also expected to drastically alter the field-effect transport properties of InAs QMs through changes in both the carrier effective mass and charge centroid position in the channel. Detailed transport studies of back-gated (50 nm $SiO_2$ gate dielectric) devices were performed, and the effective electron mobility was extracted as a function of the applied gate-voltage (i.e., vertical electric-field), temperature and QM thickness. Current flow in the conventional MOSFETs occurs near the oxide-semiconductor interface. The distance of the charge centroid from the surface ($z_D$) determines the surface scattering rate. In Si MOSFETs, $z_D$ is heavily affected by the gate-field, reducing significantly at high gate-fields, resulting in the observed field-induced degradation of carrier mobility. However, for a highly confined system, such as ultrathin InAs QMs, $z_D$ is expected to be nearly unaffected by the gate-field and is instead determined by the thickness of the layer. Thus, the scattering models, especially the field and thickness dependence of surface scattering mechanisms, are expected to be drastically different in QM FETs as compared to the conventional Si MOSFETs[27].

To explore this phenomenon, we examined electron transport properties for three cases: (i) 2-D QMs (e.g., $T_{InAs}$=8nm, Fig. 4a), (ii) quasi-2-D QMs (e.g., $T_{InAs}$=18nm, Fig. 4b), and (iii) 3-D QMs (e.g., $T_{InAs}$=48nm), corresponding to the bulk case. The electron distributions were calculated for all three cases via a Poisson-Schrodinger solver (Figs. 4c-d and Fig. S7). For $T_{InAs}$= 8 nm, the charge distribution clearly follows the ground state wave function of a quantum-well for all fields, with only the lowest sub-band being populated (Fig. S7a), and more importantly, $z_D$ is ~$T_{InAs}/2$ = 4 nm and nearly independent of the applied field (Fig. 4e). On the other hand, for the 18 nm QM, at low fields only the first sub-band, which is structurally-confined, contributes to the current, while at higher fields, a second field-confined sub-band is also populated (Fig. S7b). For the 48 nm-thick QM, transport takes place in multiple field-confined sub-bands beneath the gate oxide, with the charge centroid moving closer to the oxide interface at high fields (Fig. S7c). The final case is identical to the characteristics observed in conventional Si MOSFETs[28].

Figure 4f shows the experimental effective mobility as a function of the back-gate voltage for $T_{InAs}$= 8, 18 and 48 nm at 100 K as extracted from the transfer characteristics (Fig. S8). The dashed lines represent the theoretical mobility model considering surface roughness (SR), surface polar phonon (SPP) and impurity scattering events (Fig. S9), showing a close fit to the experiments. Although multiple scattering mechanisms were considered, surface roughness scattering was found to be dominant at low temperatures (e.g., 100 K) since phonons are mostly frozen. From the electrical measurements, the 8 nm QM shows almost no field-dependence of mobility, except at very high fields, which is in distinct contrast to conventional MOSFET model. This is well explained by the calculated charge carrier distribution profiles of Figs. 4c and 4e, where the position of charge centroid for a truly 2-D system was found to exhibit minimal field dependence. On the other hand, the 48 nm QM shows the expected bulk MOSFET behavior, with a monotonic decrease in the mobility as a function of the field. The most interesting case occurs for the 18 nm QM. At low fields, transport occurs within a single structurally-confined sub-band; at slightly higher fields, the sub-band becomes field-confined, and the mobility drops with the gate voltage; finally, at a $V_{GS}$-$V_{th}$~4V a second structurally-confined sub-band is populated. Here, the observed kink in the mobility versus field plot represents the onset of population of the 2$^{nd}$ sub-band, which is once again consistent with the calculated charge density distribution profiles. The results highlight the importance of quantization on the surface roughness scattering of QMs, and while thinner layers are more desired for scaled FET applications, the optimal thickness need to be wisely selected as a compromise with the enhanced surface scattering rates.

In conclusion, through detailed experiments and theoretical modeling, InAs QMs are presented as a model material system for elucidating the dominant role of the quantum confinement in the basic field-effect transport properties of carriers, including effective mobility and quantum resistance. InAs is specifically ideal for this study given the large Bohr radius which allows for heavy quantization at sub-20 nm thicknesses, and the ease of ohmic metal contact formation. Besides understanding the basic transport physics as a function of thickness, this work presents important practical implications for exploring the ultimate performance limits of ultrathin body InAs QM FETs, as well as providing guidelines for future device designs. Importantly, the contact resistance and mobility models in QM FETs are found to be drastically different than the previously established universal models for conventional MOSFETs, thereby, presenting a new insight into the device physics of 2-D field-effect transistors. While InAs QMs were explored here, the results are generic for any structurally quantum confined material system.

**Figure Captions**

**Figure 1.** TEM analyses of InAs QMs. (a) Low magnification and (b) high resolution TEM images of a 7 nm InAs QM capped with $SiO_2$ on a $Si/SiO_2$ substrate.

**Figure 2.** FTIR measurements of InAs QMs on $CaF_2$ substrates. (a) The transmission spectra for various InAs QM thicknesses, illustrating clear steps, corresponding to the optical transitions between the 2-D sub-bands. (b) The calculated ground and excited states for electrons and holes (light and heavy). (c) Comparison of absorption steps from FTIR and the first four theoretical interband transition energies (e1-hh1, e1-lh1, e2-hh2, e2-lh2) for various InAs QM thicknesses. Note that for 14 and 19 nm InAs QMs, e1-hh1 and e1-lh1 transitions are so close that cannot be clearly resolved by FTIR.

**Figure 3.** Quantum contact resistance of InAs QMs. (a) Typical $I_{DS}$-$V_{GS}$ results of a 18 nm-thick InAs QM with different channel lengths, $L$. (b) ON-resistance of InAs QMs ($T_{InAs}$, 5-48 nm) at $V_{GS}$-$V_{th}$=9 V as a function of the channel length. (c) Experimentally extracted (red symbols), NEGF simulated (dashed line), and analytically calculated (dotted line) contact resistance values versus thickness. The error bars in the experimental data corresponds to the standard error or regression. (d) Calculated energy band diagrams of the contact interface for 8 nm (top) and 18 nm (bottom) thick InAs QMs. The Fermi level, $E_f$, conduction band edge, $E_c$, and ground and excited states (e1- e4) are labeled. The sub-bands that are below $E_f$ are ohmically contacted while those that are above are Schottky contacted.

**Figure 4.** Electron transport analyses of back-gated InAs QM FETs. (a) (b) The calculated band diagrams of back-gated InAs QM FETs for $T_{InAs}$ = 8 nm and 18 nm, respectively. The calculation

is shown for both low (left) and high (right) vertical gate fields. The population percentage for each sub-band is also labeled. (c) (d) The calculated carrier density vs. position as a function of the applied gate field for $T_{InAs}$ = 8 nm and 18 nm, respectively. (e) The calculated charge centroid position as a function of the gate-field for different InAs QM thicknesses. f, Experimental (solid lines) and simulated (dashed lines) effective mobilities as a function of gate voltage for varied InAs QM thicknesses at 100 K and $V_{DS}$=10 mV.

**Acknowledgements**

This work was funded by FCRP/MSD Focus Center and NSF E3S Center. The materials characterization part of this work was partially supported by the Director, Office of Science, Office of Basic Energy Sciences, and Division of Materials Sciences and Engineering of the U.S. Department of Energy under Contract No. De-Ac02-05Ch11231, and the Electronic Materials (E-Mat) program. The Advanced Light Source is supported by the Director, Office of Science, Office of Basic Energy Sciences, of the U.S. Department of Energy under Contract No. DE-AC02-05CH11231. A.J. acknowledges a Sloan Research Fellowship, NSF CAREER Award, and support from the World Class University program at Sunchon National University. Y.-L.C. acknowledges support from the National Science Council, Taiwan, through grant no. NSC 98-2112-M-007-025-MY3. R.K. and M.M. acknowledge an NSF Graduate Fellowship and a postdoctoral fellowship from the Danish Research Council for Technology and Production Sciences, respectively. J.G. acknowledges support form NSF and SRC.

**Supporting information**

Sample preparation and FTIR measurement details; temperature dependent FTIR measurement; calculated 2-D sub-band edges; quantum resistance calculations; NEGF simulations of contact resistance; simulations of the charge density profiles; low-temperature transfer characteristics of InAs QM FETs; theoretical mobility models; temperature, vertical-field, and thickness dependency of the effective mobility. This material is available free of charge via the Internet at http://pubs.acs.org.

**Figure 1**

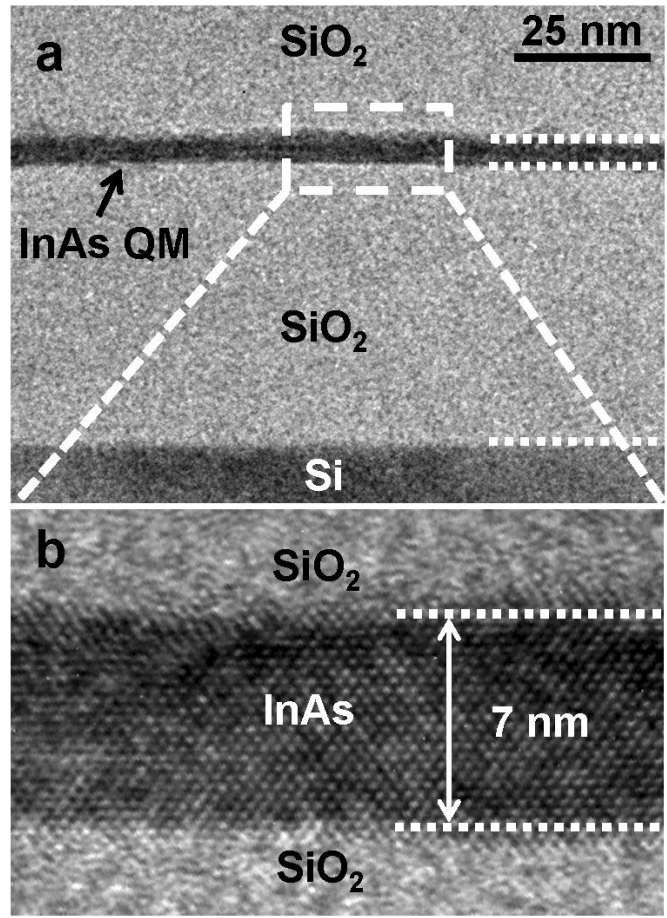

**Figure 2**

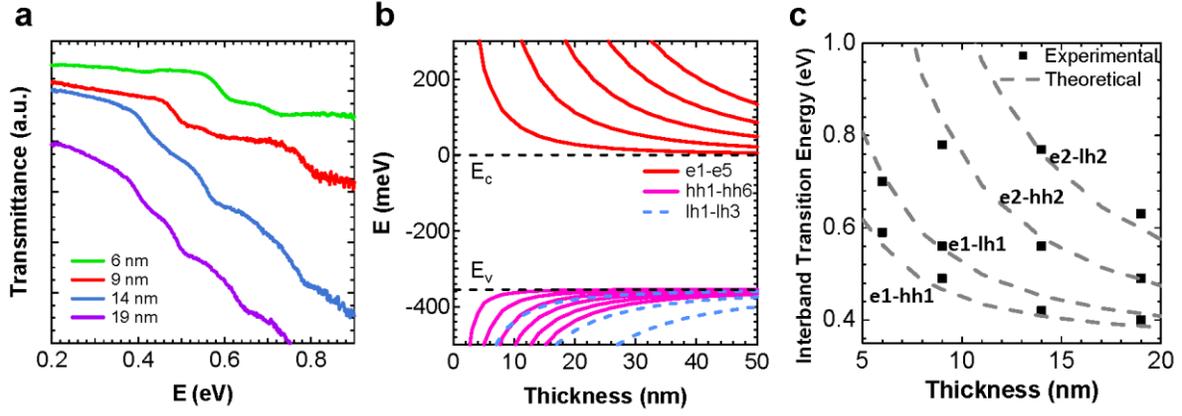

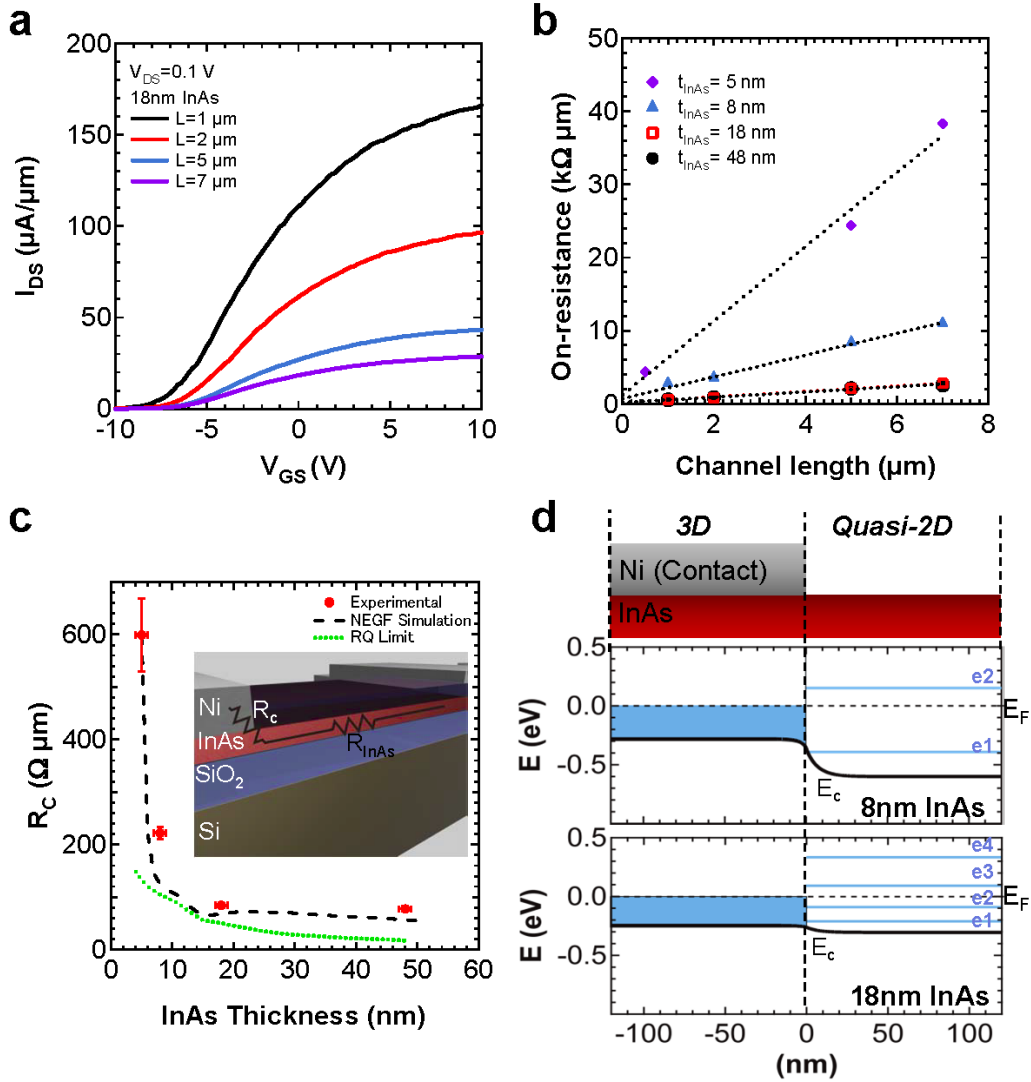

**Figure 4**

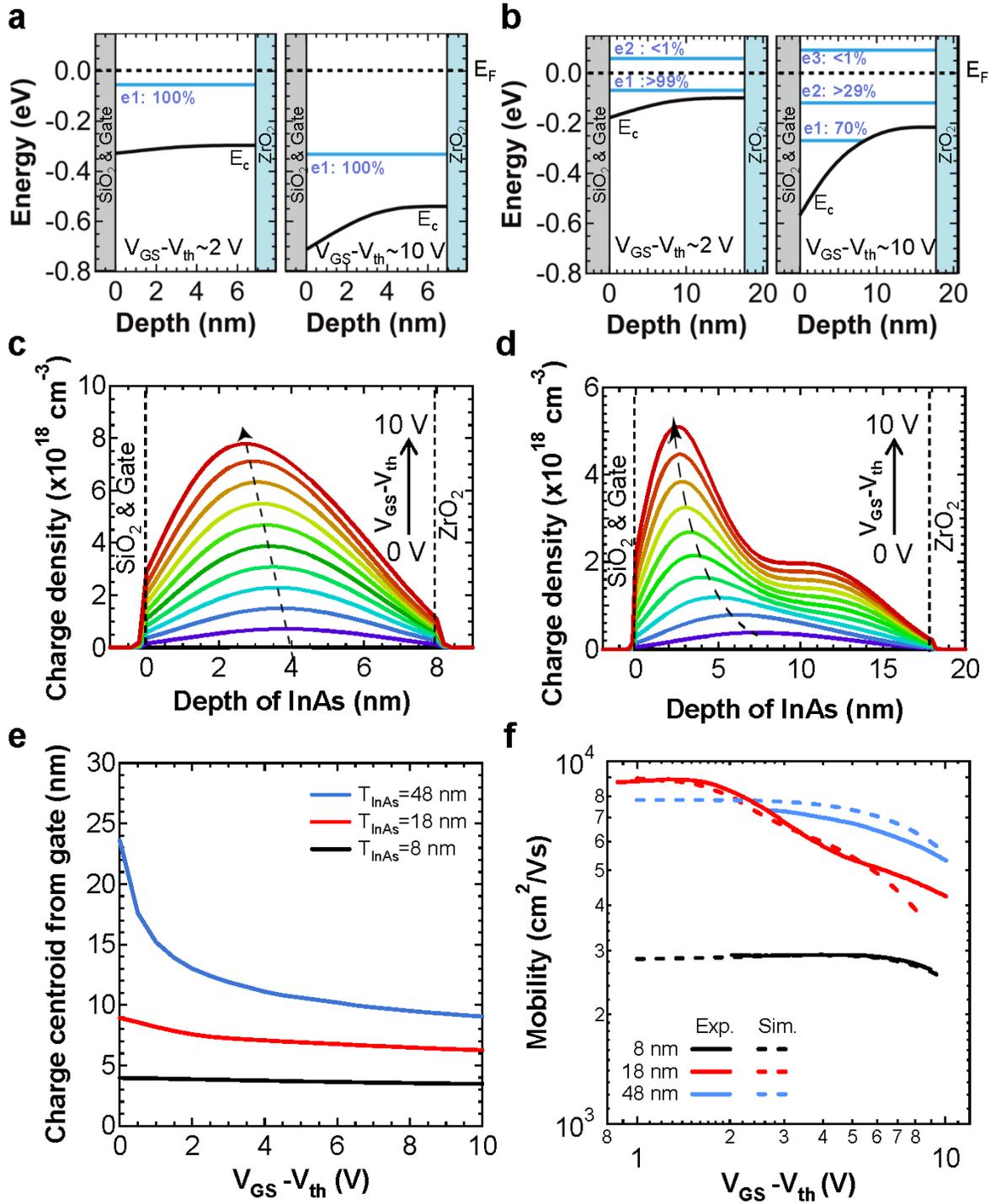

# TOC Figure

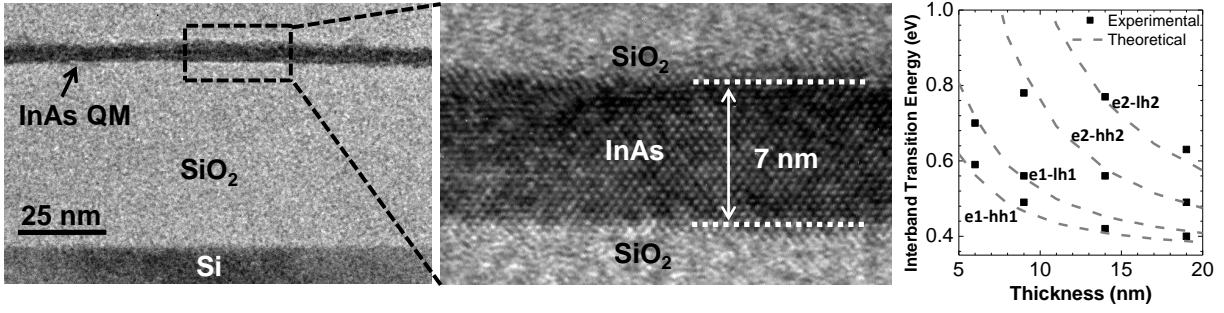

# SUPPORTING INFORMATION

## Highly Quantum-Confined InAs Nanoscale Membranes

Kuniharu Takei[1,2,3,†], Hui Fang[1,2,3,†], Bala Kumar[4,†], Rehan Kapadia[1,2,3], Qun Gao[4], Morten Madsen[1,2,3], Ha Sul Kim[1,2,3], Chin-Hung Liu[5], Yu-Lun Chueh[5], Elena Plis[6], Sanjay Krishna[6], Hans A. Bechtel[7], Jing Guo[4], Ali Javey[1,2,3,]∗

[1]Electrical Engineering and Computer Sciences, University of California, Berkeley, CA, 94720.

[2]Materials Sciences Division, Lawrence Berkeley National Laboratory, Berkeley, CA 94720.

[3]Berkeley Sensor and Actuator Center, University of California, Berkeley, CA, 94720.

[4]Electrical and Computer Engineering, University of Florida, Gainsville, FL, 32611

[5]Materials Science and Engineering, National Tsing Hua University, Hsinchu 30013, Taiwan, R. O. C.

[6]Center for High Technology Materials, University of New Mexico, Albuquerque, NM 87106.

[7]Advanced Light Source, Lawrence Berkeley National Laboratory, Berkeley, CA 94720

[†] These authors contributed equally to this work.

∗ Correspondence should be addressed to A.J. (ajavey@eecs.berkeley.edu).

**Sample preparation**

**Epitaxial layer transfer:** AlGaSb (60 nm) and InAs heteroepitaxial thin films were grown by molecular beam epitaxy on a GaSb handling wafer. The InAs layer was pattern etched into nanoribbons with a Polymethylmethacrylate (PMMA) mask (pitch, ~840 nm; line-width, ~350 nm) using a mixture of citric acid (1 g per ml of water) and hydrogen peroxide (30%) at 1:20 volume ratio (etch rate, ~1 nm/sec). To release the InAs nanoribbons from the source substrate, the AlGaSb sacrificial layer was selectively etched by ammonium hydroxide (3% in water) solution for 110 min prior to PMMA stripping in acetone. Next, an elastomeric polydimethylsiloxane (PDMS) substrate (~2 mm thick) was used to detach the nearly full released InAs ribbons from the GaSb donor substrate and transfer them onto a Si/SiO$_2$ receiver substrate. During the transfer process, any strain in the InAs ultrathin layers due to a lattice mismatch with the growth substrate is released[1], resulting in free-standing, fully-relaxed InAs QMs on insulator. For FTIR studies, the InAs layer was pattern etched and transferred with a resist mask (pitch, ~10 µm; line-width, ~5 µm). The resist was stripped after transferring the InAs layers onto CaF$_2$ substrates. InAs membranes transferred without a resist cap are often thinner by ~1 nm than those with a cap as observed from TEM studies.

**Device fabrication:** InAs layers were transferred onto thermally grown 50 nm-thick SiO$_2$ on *p+* Si substrates using the ELT process. Next, Ni/Au metal S/D contacts were fabricated by lithography, evaporation and lift-off. Contact annealing was carried out in N$_2$ ambient at 300°C for 1 min. Subsequently, a 7 nm thick ZrO$_2$ layer was deposited as a capping layer using atomic layer deposition at 130°C with tetrakis(ethylmethylamido)ziroconium and water as the precursors. Finally, via contacts were opened on the bonding pads by 2% HF etch. The electrical measurements were performed using the *p+* Si substrate as the global back gate.

**FTIR measurements**

The optical transmittance data of InAs QMs ($T_{InAs}$=5-48 nm) were measured by FTIR micro-spectroscopy using a Thermo Scientific Nicolet Continuμm Infrared Microscope and a Thermo Scientific Nicolet 6700 FTIR spectrometer equipped with a $CaF_2$ beamsplitter at Beamlines 1.4 and 5.4 at the Advanced Light Source. InAs layers were transferred to $CaF_2$ crystal substrates (International Crystal Laboratories, double side polished, 1mm thick) using the ELT technique. Spectra in the 1200-7700 cm$^{-1}$ (0.15 eV-0.95 eV) range were collected with an aperture size of 50×50 μm$^2$ and 8 cm$^{-1}$ (1 meV) resolution and averaged 512 times.

**FTIR spectrum of a 49 nm thick InAs membrane**

Figure S1 shows the transmittance spectrum of a 49 nm-thick InAs layer transferred onto a CaF$_2$ substrate, exhibiting no clear optical transition steps. As expected, this shows that the 49 nm InAs membrane behaves *nearly* like a 3-D material, in terms of absorption and electronic band structure.

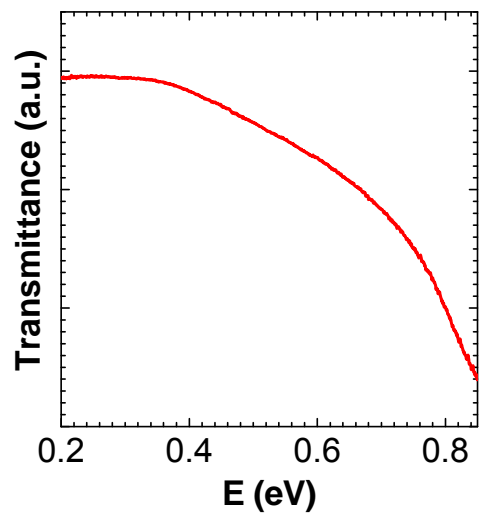

**Figure S1.** The FTIR transmission spectrum for a 49 nm thick InAs membrane.

**Extraction of absorption onsets from FTIR spectra of InAs QMs**

Due to the band edge roughness, the absorption steps observed in Fig. 2a are not strictly sharp. To determine the absorption onsets from the transmittance spectra, the first derivatives were obtained, where the location of each peak (downward) corresponds to the onset energy of an interband transition. Firstly, the transmittance spectra are smoothed by the Savitzky-Golay function with second polynomial order and 51 points, and then first order derivatives were calculated, as shown in Fig. S2.

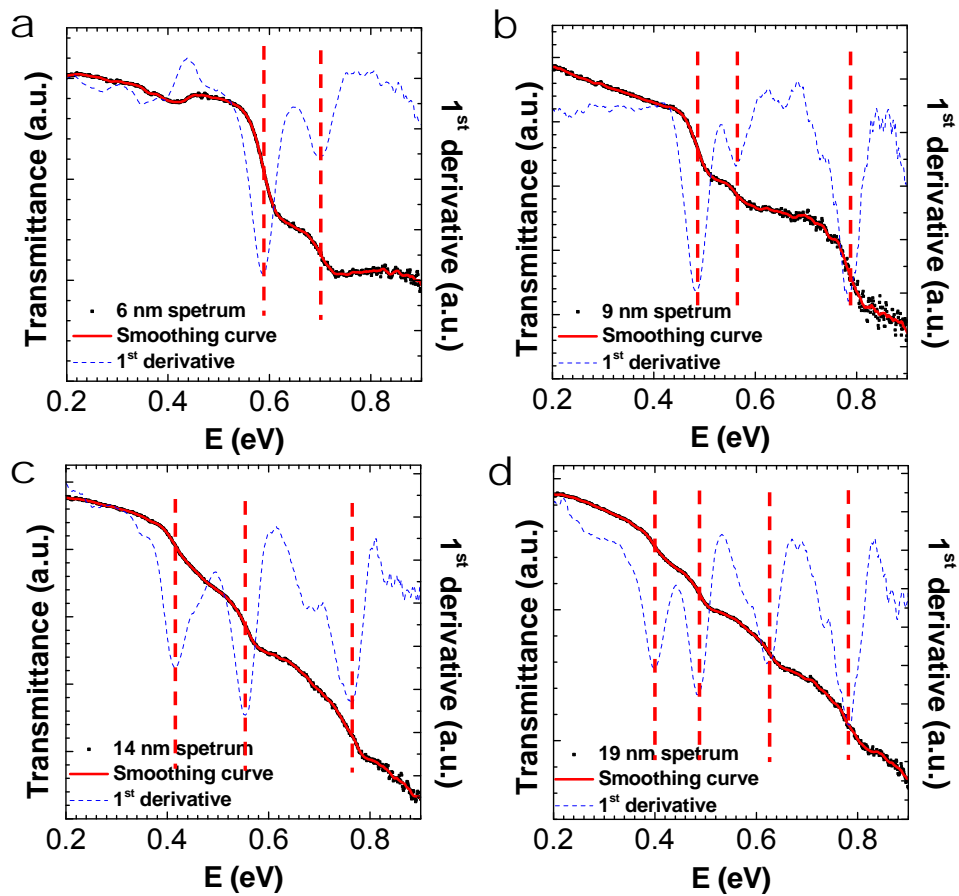

**Figure S2.** Absorption onsets determination for the FTIR transmission spectra of (a) 6 nm, (b) 9 nm, (c) 14 nm and (d) 19 nm InAs QMs.

**Low-temperature FTIR**

Figure S3 shows the comparison of the FTIR spectra of a 6 nm thick InAs QM at 4 K and 300 K. Using the same approach to determine the transition onsets described above, a clear blue shift of ~0.056 eV in the interband transition energies is obtained from 300 K to 4 K. The magnitude of this shift can be explained by the analytical equation: $E=E_g+E_{hn}+E_{en'}$, where the temperature dependence of InAs $E_g$ is given by the following empirical equation,

$$E_g(T) = 0.415 - 2.76 \times 10^{-4} T^2 /(T+83)(eV) \text{ [Ref. 2]}$$

Assuming that there is no temperature dependence in $E_{hn}$ and $E_{en'}$, then for any given interband transition:

$$\Delta E_{300K \to 4K} = \Delta E_{g300K \to 4K} = E_g(4K) - E_g(300K) = 0.064 \text{ eV}$$

This calculation is consistent with the observed blue shift, once again depicting that the steps in the absorption spectra arise from the quantized electronic band structure of InAs QMs.

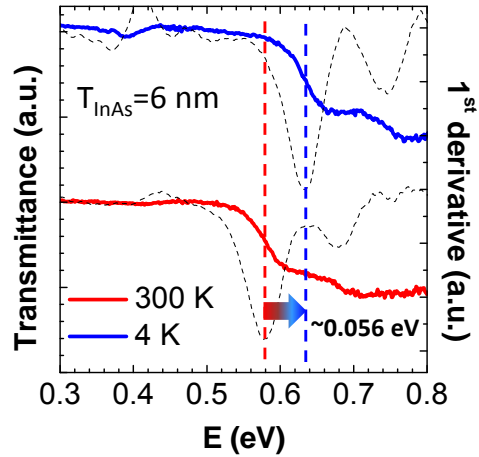

**Figure S3.** The FTIR transmission spectra of a 6 nm thick InAs QM at 300 K (red) and 4 K (blue).

## Calculated 2-D sub-band edges

By modeling the InAs QM as a finite potential well structure, the n'th sub-band edge energy can be numerically calculated from 1-D time independent Schrodinger equation:

$$-\frac{\hbar^2}{2m^*}\frac{d^2\psi(z)}{dz^2}+V(z)\psi(z)=E\psi(z),$$

where $\hbar$ is the reduced Planck's constant, $m^*$ is the effective mass, $\psi(z)$ is the wavefunction, $V(z)$ is the potential energy, $E$ is the eigenenergy due to confinement.

The potential barrier height of InAs QMs is determined by the difference of the electron affinity across the interface. Bulk effective masses of electrons (e), heavy holes (hh) and light holes (lh) were used here with $m_e^* = 0.023\ m_0$, $m_{hh}^* = 0.41\ m_0$ and $m_{lh}^* = 0.026\ m_0$, where $m_0$ is the electron rest mass in vacuum. Note that the inaccuracy due to the effective mass change from confinement[3] is very small in the range explored in this study. For example, for 6 nm-thick InAs, the energy for e1 changes from 0.18 eV to 0.17 eV if $m_e^*$ is increased by 50%.

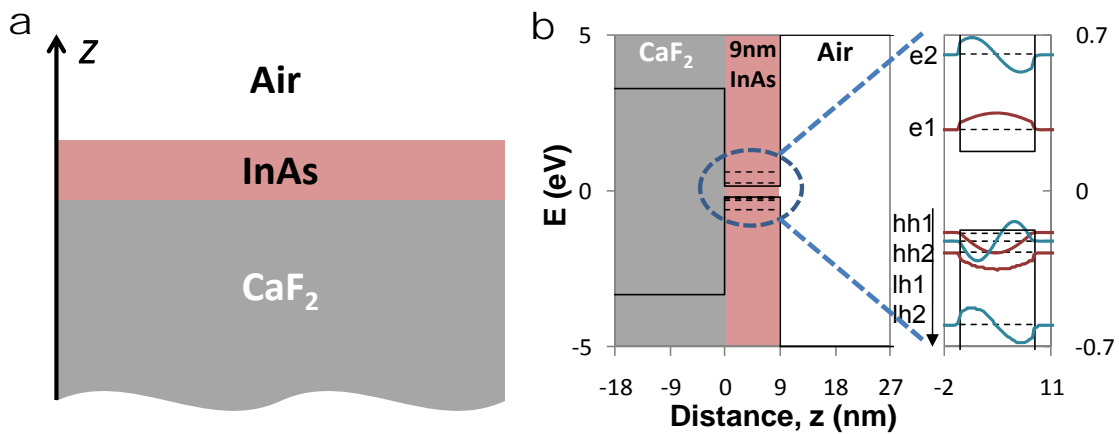

**Figure S4.** (a) The schematic of an InAs QM with the boundaries of air and CaF$_2$ used for the sub-band edge calculations, corresponding to the experimental structure used for FTIR studies. (b) The potential energy for electrons and holes as a function of the distance in z direction for $T_{InAs}$=9 nm, showing a confined quantum well for electrons and holes. The ground and the first excited states are also shown.

**Quantum resistance calculations**

The number of populated 1-D sub-bands is a function of the Fermi energy, and thus, doping level. Figure S5B shows the calculated $R_Q$ as a function of Fermi energy for various thicknesses[4]. Critically, a 2-D material behaves differently from a 1-D material, where the conductance is constant between 1-D sub-bands, since within a single 2-D sub-band, the number of transport modes increases by increasing the Fermi energy, as shown in Fig. S5a. This enables greater control over contact resistance in a 2-D material compared to a 1-D material. The $R_Q$ in Figure 3c is calculated by considering the $E_F$ corresponding to a doping concentration of $3\times10^{18}$ cm$^{-3}$ (which is the surface/interface electron concentration previously reported for bulk InAs)[5,6], and then utilizing that $E_F$ to calculate the number of available transport modes at T= 0 K. It should be noted that this assumes that both the probability of an electron being injected from the source and the injection from the mode into the drain is unity.

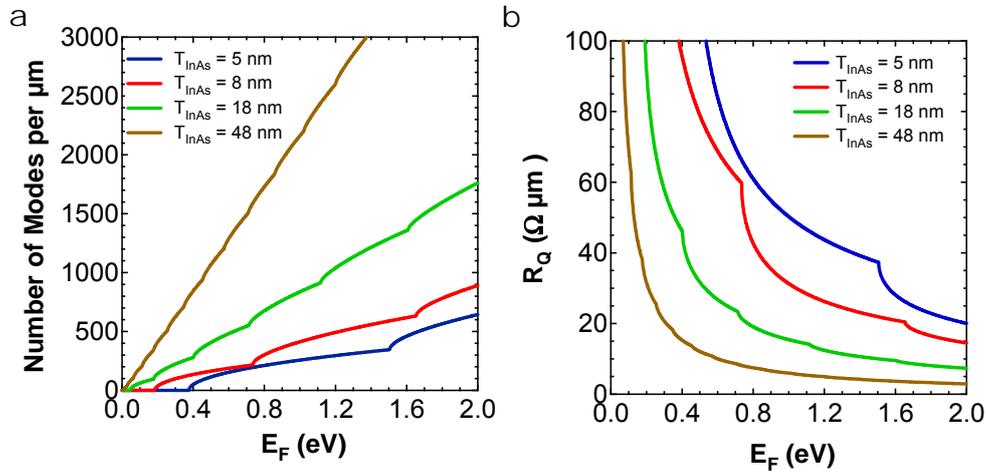

**Figure S5.** (a) Calculated number of 1D transport modes in InAs QMs of indicated thicknesses as a function of the Fermi level position in respect to the conduction band edge (i.e., doping concentration). The contribution of a new sub-band to transport is evident as a kink in the plots as the Fermi level in increased. (b) Calculated quantum resistance, assuming the transmission probability through each mode is 1.

**NEGF simulation of contact resistance**

The non-equilibrium Green's function (NEGF) method is employed to perform theoretical modeling of the contacts, and the approach is described in Ref. 7. The simulation describes two-dimensional quantum transport from the contact to the XOI body. The conductance per width ($G/W=1/R_c$) is given as:

$$\frac{G}{W} = 2\frac{e^2}{h} \int_{-\infty}^{+\infty} dE \cdot \sum_{k_z} Tr(E)(-\partial f(E + \hbar^2 k_z^2 / m_z)/\partial E)$$

where $f$ is the Fermi distribution function, $h$ is the Plank constant, $Tr(E)$ is the transimission function and $m_z$ is the effective mass along width. Since transmission probability $T(E)$ and $H$ are independent of $k_z$, $G/W$ can be explained as,

$$\frac{G}{W} = 2\frac{e^2}{h} \int_{-\infty}^{+\infty} dE \cdot T(E) \cdot (-\partial f_{1D}(E)/\partial E)$$

Here, $f_{1D}(E)$ is the 1D $k$-summed Fermi function described as,

$$f_{1D}(E) = \left(\frac{m_z k_B T}{2\pi \hbar^2}\right) F_{-1/2}\left(-\frac{E}{k_B T}\right)$$

where $F_{-1/2}$ is the Fermi-Dirac integral of order -1/2.

Here, a doping concentration of $3\times10^{18}$ cm$^{-3}$ and an Schottky barrier height of -0.15 eV to the conduction band edge are used, corresponding to the values reported in literature[5,6]. Ballistic transport is assumed, which corresponds to the limit of zero source-channel XOI diffusion resistance. In the simulation, the contact length is chosen to be $L_C$ =280 nm because further increase of $L_C$ plays a negligible role on the simulated contact resistance. This choice most accurately reflects the actual structure while minimizing the computation time.

## Quantum simulations of the charge density profiles

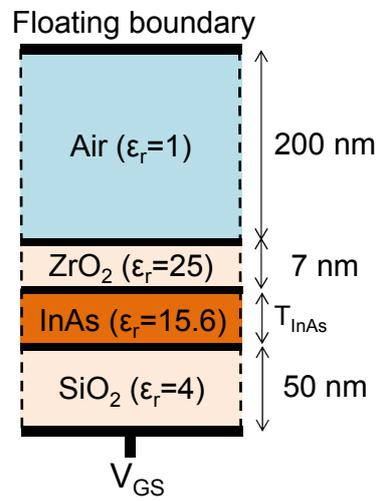

**Figure S6.** Simulated structure for the charge density profiles. Effective masses for $SiO_2$ and $ZrO_2$ used here are $0.5m_0$ and $0.3m_0$, respectively[8].

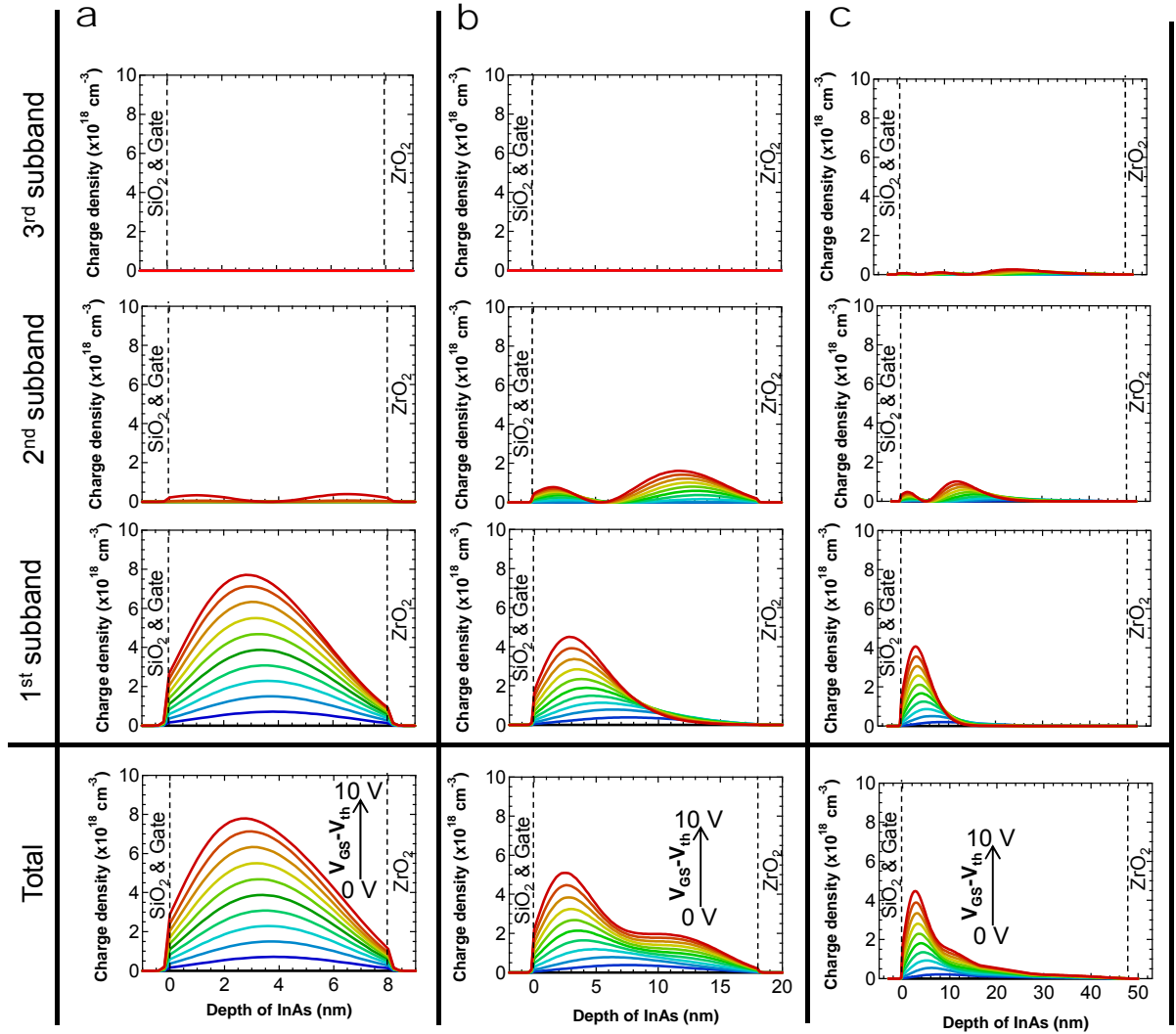

**Figure S7.** Calculated carrier density profiles, showing the population density of $1^{st}$ – $3^{rd}$ sub-bands as a function of the vertical-field bias ($V_{GS} - V_{th}$) for (a) 8 nm, (b) 18 nm and (c) 48 nm InAs QMs. The devices are back-gated with 50 nm $SiO_2$ as the gate dielectric.

**Low-temperature transfer characteristics of back-gated QM FETs**

Figure S8 shows representative $I_{DS}$-$V_{GS}$ curves for 8 nm, 18 nm and 48 nm thick InAs XOI FETs at 100 K. For 18 nm InAs, a kink in the $I_{DS}$ at $V_{GS}$~2 V (the position is marked with an arrow in Fig. S8a) indicates the onset of population for the 2$^{nd}$ sub-band. Effective mobility plots in Fig. 4f were extracted from the $I_{DS}$-$V_{GS}$ curves of Fig. S8.

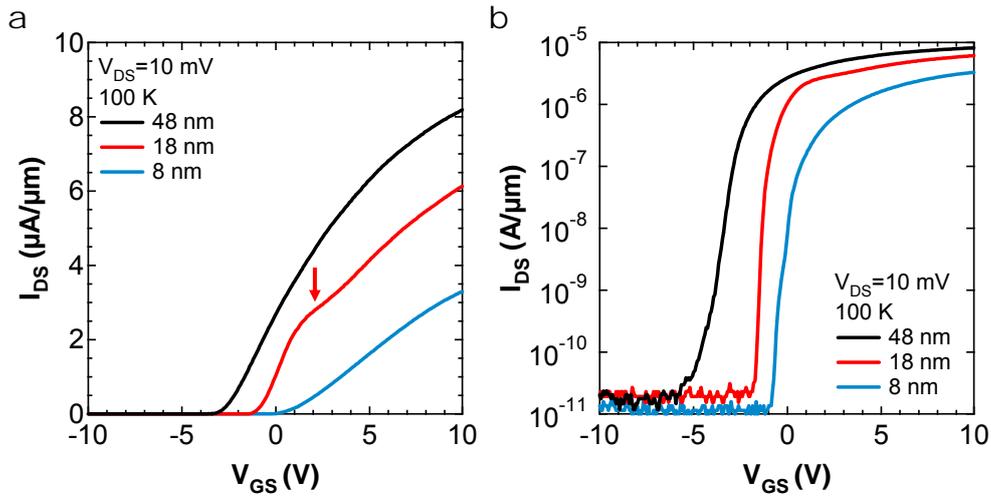

**Figure S8.** (a) and (b) Transfer characteristics of InAs QMs with 8 nm, 18 nm and 48 nm thicknesses at $V_{DS}$=10 mV and 100 K sample temperature.

**Theoretical mobility models**

In order to model the extracted effective mobility, multiple scattering mechanisms were considered, as listed below. The overall scattering rate was given by the sum of the individual scattering rates as:

$$\tau^{-1} = \sum_i \tau_i^{-1}$$

The mobility was then calculated in the momentum relaxation approximation as:

$$\mu = \frac{e\tau}{m^*}$$

The scattering rates considered were:

(i) Surface polar optical phonon scattering [9,10]

$$\frac{1}{\tau_{SPP}} = \frac{2e^2 F^2 m^*}{\pi \hbar^3 \varepsilon_{ox}} N_{SPP}^{\pm} \int_0^{2\pi} \frac{e^{-2\varepsilon_{ox} P_{\mp}(\theta) z_0} P_{\mp}(\theta)}{(P_{\mp}(\theta) + \lambda_{SO})^2} d\theta$$

$$F^2 = \frac{\hbar \omega_{SO}}{2\,\varepsilon_0} \left( \frac{1}{\varepsilon_{ox}^0 + 1} - \frac{1}{\varepsilon_{ox}^\infty + 1} \right)$$

$$P_{\mp}(\theta) = \sqrt{\frac{2E_i}{\hbar \omega_{SPP}} \mp 1 - 2\sqrt{\left(\frac{E_i}{\hbar \omega_{SPP}} \mp 1\right) \frac{E_i}{\hbar \omega_{SPP}}} \cos(\theta)}$$

$\varepsilon_{ox}^0$: low-frequency SiO$_2$ dielectric permittivity

$\omega_{SO}$: Surface optical phonon angular frequency

$\lambda_{SO}$: Surface optical phonon wavelength

$N_{SPP}$: Surface polar phonon occupation statistics

$\omega_{SPP}$: Surface polar phonon angular frequency

$E_i$ = electron energy

(ii) Surface roughness scattering [11]

$$\frac{1}{\tau_{SR}} = \frac{e^2 m^* |\int \Psi_v \left[\frac{\Delta V_m}{\Delta_m}\right] \Psi_u dz|^2 \Delta_m^2 L^2}{2\hbar^3} \int_0^{2\pi} \frac{d\theta}{\epsilon(q)(1 + L^2 q^2/2)^{\frac{3}{2}}}$$

$$\epsilon(q) = 1 + \frac{e^2 m^*}{2\varepsilon_{InAs} q \pi \hbar^2} F(q)$$

$$F(q) = \sum_m \int dz \int dz' |\Psi_m(z)|^2 |\Psi_m(z')|^2 e^{q|z-z'|}$$

$\Psi_v$: envelope function for $v^{th}$ subband
$\Delta V_m$: surface roughness perturbing potential
$\Delta_m$: RMS surface roughness
L: autocorrelation length

(iii) Interface charge scattering [12,13,14]

$$\mu_{im} = \alpha(V_G - V_T)^{\frac{3}{2}} + \mu_0(t)$$

$\mu_0$: baseline interface roughness mobility

$\alpha$: gate field fitting parameter

The main components considered were (i) surface polar phonons (SPP), which enabled modeling of the temperature dependence; (ii) surface roughness scattering (SR), which enabled modeling of the gate dependence; and finally (iii) an empirical impurity scattering term, which accounted for the scattering not represented via the SR and SPP models. Critically, the gate dependence (equivalently, sheet charge density) of mobility can be explained by the surface roughness scattering. This scattering mechanism is closely tied to the envelope functions, or charge distributions, of each sub-band, as evident from the sub-band wavefunction overlap integral in the analytical equation for SR scattering above. Due to this dependence, the charge distribution as well as the number of sub-bands populated heavily affect the gate dependence of the SR mobility. There is a visible 'kink' in the effective mobility versus sheet charge density plot for the 18 nm QM (Fig. 4f and S9), which coincides with the population of the second sub-band. The change in slope, causing the kink, can be understood by noting that the charge centroid of the $2^{nd}$ sub-band is not a function of sheet density. Thus the change in charge centroid as a function of gate field, or equivalently, sheet density is significantly reduced.

**Temperature, vertical-field, and thickness dependency of the effective mobility**

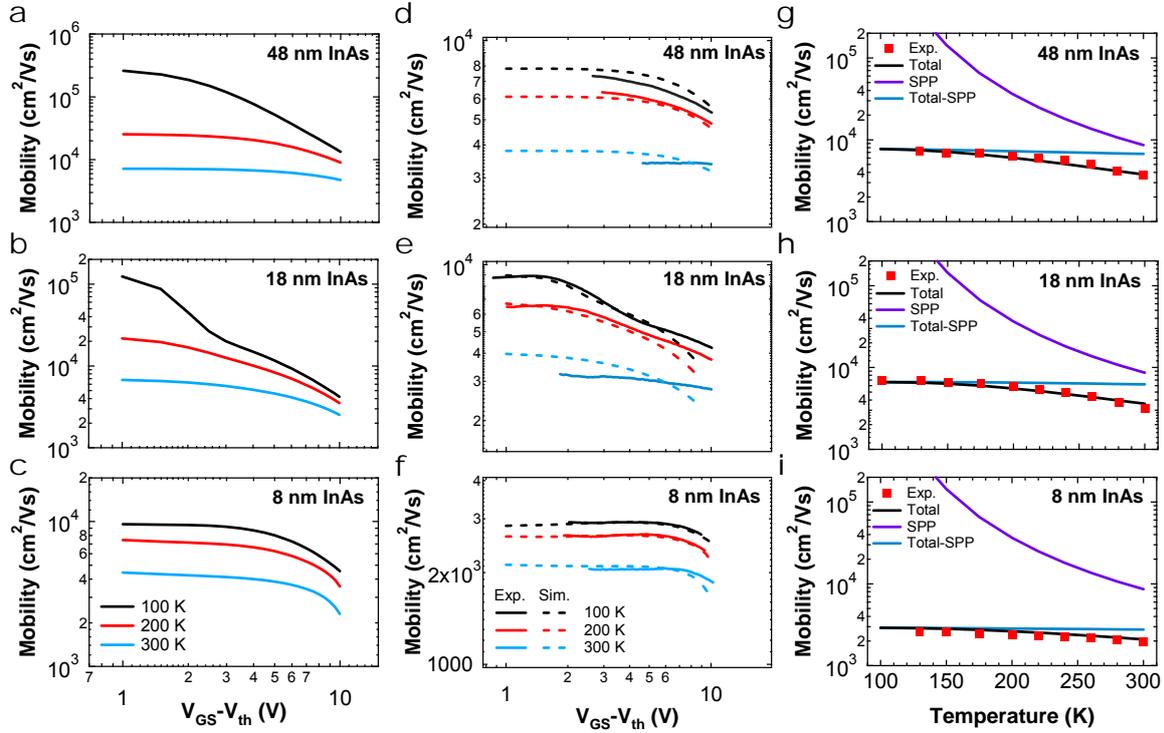

**Figure S9.** (a), (b), (c) InAs QM mobility calculated by only including the surface roughness (SR) and surface polar phonon (SPP) scattering events, which is in qualitative agreement for the experimental data. The mobility for 8nm QM is approximately independent of the gate-field because the charge centroid remains near the middle of the body regardless of the gate voltage, which is different from a thicker body in which the applied gate voltage moves the centroid closer to the surface (Fig. 4c&d, Fig. S7), resulting in a decrease of mobility due to enhanced SR scattering. For the 8nm QM, only one sub-band is occupied and an ideal 2D electron gas is obtained, which is in contrast to the thicker body case, where multiple sub-bands are populated. For a body thickness of 18nm QM, the shoulder in the mobility curve at $T$=100K and $V_{DS}$=10 mV as evident in both the simulation and experimental data is due to the second sub-band being populated. Though qualitative features are understood by the interplay of quantum confinement effects and SR scattering, quantitative difference exists, likely due to other scattering mechanisms such as dislocations, ionized impurities, and surface charge traps. The exact scattering mechanism responsible for the difference requires further study. However, by superimposing the mobility in the first column (SR and SPP) with an empirical model for dislocation scattering, quantitative agreement with experimental data can be achieved, as shown in (d), (e), (f). To determine the extent the surface polar phonons determine the temperature dependence of mobility at $V_{GS}$-$V_{th}$=9 V and $V_{DS}$=10 mV, the SPP contribution to mobility (purple line) was subtracted from the total calculated mobility (black line) to produce the difference

(blue line) as shown in (g), (h), (i). This illustrates how nearly the entire temperature dependence can be ascribed to SPP.

The temperature dependency of the mobility was studied in detail for various thicknesses in order to elucidate the phonon scattering events. For all three thicknesses, the experimentally obtained effective mobility is found to increase as the temperature is reduced from 300K to ~150K. The temperature dependency of mobility weakens as the temperature is reduced more, with the mobility eventually saturating and becoming nearly independent of temperature for $T<\sim150K$. A close fit between experiment and simulation was obtained (Fig. S9). As evident in Fig. S9 (g-i), the temperature dependence of the mobility is mainly due to the surface polar phonon (SPP) scattering, which increases with increasing temperature.